\documentstyle[preprint,aps,prl,psfig]{revtex}
\begin{document}
\newcommand{\sinf}{\raisebox{-.7ex}{$\stackrel{<}{\sim}$}}
\newcommand{\ssup}{\raisebox{-.7ex}{$\stackrel{>}{\sim}$}}

\ifx\undefined\psfig\def\psfig#1{    }\else\fi
\ifpreprintsty\else
\twocolumn[\hsize\textwidth
\columnwidth\hsize\csname@twocolumnfalse\endcsname       \fi    \draft
\preprint{  } \title  {Exciton-exciton interaction engineering in
coupled GaN quantum dots}
\author  {Sergio De Rinaldis$^{1,2}$, Irene D'Amico$^{1,3}$, and Fausto Rossi$^{1,3,4}$}
\address{
$^1$ Istituto Nazionale per la Fisica della Materia (INFM) \\
$^2$ National Nanotechnology Laboratory (NNL) and
Istituto Superiore Universitario
per la Formazione Interdisciplinare (ISUFI),
73100, Lecce, Italy;\\
$^3$ Institute for Scientific Interchange (ISI),
10133 Torino, Italy \\
 $^4$ Dipartimento di Fisica, Politecnico di Torino,
10129 Torino, Italy
}
\date{\today} \maketitle

\begin{abstract}

We present a fully three-dimensional study of the multiexciton optical response of vertically coupled GaN-based quantum dots via a direct-diagonalization approach. The proposed analysis is crucial in understanding the fundamental properties of
few-particle/exciton interactions and, more important, may play an essential role in the design/optimization of semiconductor-based quantum information processing schemes. In particular, we focus on the interdot exciton-exciton coupling, key ingredient in recently proposed all-optical quantum processors. Our analysis demonstrates that there is a large
window of realistic parameters for which both biexcitonic shift and oscillator strength are compatible with such implementation schemes.

\end{abstract}

\ifpreprintsty\else\vskip1pc]\fi \narrowtext


Semiconductor quantum dots (QDs) are systems of paramount interest in 
nanoscience and nanotechnology\cite{dot}.
They are the natural evolution of band-gap/wavefunction engineering in
semiconductors; however, in contrast to
quantum wells and wires, they exhibit a discrete ---i.e.,
atomic-like--- energy spectrum and, more important, their optical response is dominated by
few-particle/exciton effects.
Moreover several QD-based technological applications have been proposed, such as
lasers\cite{Saito}, charge-storage devices\cite{Finley}, fluorescent
biological markers\cite{Bruchez}, and all-optical quantum information
processors\cite{Biolatti1}.

So far, self-organized QDs have been successfully fabricated using a wide
range of semiconductor materials; they include III-V QD structures based
on GaAs as well as on GaN compounds. GaAs-based QDs have been well characterized
and their electronic structures have been widely 
studied\cite{Rinaldi,Biolatti2},
whereas the characterization of GaN-based systems is still somewhat fragmentary
and their electronic properties have been studied only 
recently\cite{Andreev}.
GaAs- and GaN-based nanostructures exhibit very different
properties: GaN systems have a wider bandgap ($3.5$\,eV) compared to
GaAs-based ones ($1.5$\,eV). Moreover, whereas GaAs and most of the other
III-V compounds have a cubic (zincblende) structure, GaN (as well as other nitrides) has a hexagonal (wurzite) structure which leads to strong
built-in piezoelectric fields (of the order of MV/cm). As a consequence of
such built-in fields, in these nanostructures excitonic
transitions are red-shifted, and the
corresponding interband emission is fractions of eV below the bulk GaN band-gap.

In this Letter we shall provide a detailed investigation of the interplay between
single-particle carrier confinement and two-body Coulomb interactions in coupled GaN-based QDs. In particular, we shall analyze exciton-exciton dipole coupling versus oscillator strength:
we demonstrate that it is possible to tailor and control such non-trivial Coulomb interactions
 by varying the QD geometry (e.g., base and height), since this  in turn modifies the  wavefunctions of electrons and holes confined into the QDs as well as
intrinsic electric fields;
at the same time, our investigation shows that the oscillator strength of the ground-state exciton decreases super-exponentially with increasing QD height.

The relevance of our analysis is twofold: (i) we address a distinguished few-particle phenomenon typical of nitride QDs, i.e., the presence of an intrinsic exciton-exciton dipole coupling induced by built-in polarization fields; (ii) we provide detailed information on the set of parameters needed for the experimental realization of the quantum information processing strategy proposed in \onlinecite{DeRinaldis}.

More specifically, in our analysis we consider the range of GaN/AlN QDs presented in
\onlinecite{Arley}: the dot height will vary from $2$ to $4$\,nm and the QD-base diameter from $10$ to $17$\,nm, assuming a linear dependence between these
two parameters in agreement with experimental and theoretical 
findings\cite{Andreev}.

As already underlined,
the peculiarity of wurzite GaN heterostructures is the strong
built-in electric field which is the sum of the spontaneous
polarization and the piezoelectric field. Spontaneous polarization
charge accumulates at the GaN/AlN interfaces as a consequence  of a slight
distortion of GaN and AlN unit cells,
compared to those of an ideal hexagonal crystal. Piezoelectric fields are caused
by uniform strain along the (0001) direction.
Contrary to GaN/AlGaN quantum wells ---where the
spontaneous-polarization contribution is dominant~\cite{Cingolani}---  in QDs the
strain-induced piezoelectric field and the spontaneous-polarization potential are
of similar magnitude and sign, both oriented along the growth
direction. The strength of the
intrinsic field along such
direction is almost the same inside and outside the dot, but it is
opposite in sign.
The built-in electric field in GaN QDs and AlN barriers is calculated
according to\cite{Cingolani}:

\begin{equation}
F_{d} =
{L_{br}(P_{tot}^{br}-P_{tot}^{d})\over\epsilon_{0}(L_{d}\epsilon_{br}+L_{br}\epsilon_{d})}\ ,
\label{region1}
\end{equation}

where $ \epsilon_{br,(d)}$  is the relative dielectric constant
of the barrier (of the
quantum dot), $P_{tot}^{br,(d)}$  is the total polarization
of the barrier (of the
quantum dot), and
$L_{br,(d)}$ is the
width of the
barrier (the height of the dot).
The value of the field in the barrier $F_{br}$   is obtained by
exchanging the
indices {\it br} and {\it d}.
Equation (\ref{region1}) is derived for an alternating sequence of
quantum wells and barriers, but it is a good
approximation also in the case of   an array of
similar QDs  in the growth (z) direction.
The lateral shape of the QD is simply approximated by a bidimensional
parabolic
potential which mimics the strong in-plane carrier confinement
caused by the built-in electric field and preserves the spherical symmetry of
the ground state\cite{Andreev}. Our approach
is supported by the agreement  with the
experimental findings in \onlinecite{Andreev}.
The polarization is the sum of the spontaneous polarization charge that
accumulates at GaN/AlN interfaces and the piezoelectric one. All the
parameters are taken from Ref.~\onlinecite{Cingolani} (adapted for the case
x=1 for Al percentage in the barrier).

The above theoretical scheme has
been applied to realistic
state-of-the-art GaN QDs.  The difference between the well width of two neighboring QDs is assumed to be 8\%
  to allow energy-selective generation of ground-state excitons in neighboring QDs. The barrier width is such
to prevent single-particle tunneling and to allow at the same time
significant dipole-dipole Coulomb coupling:
the giant internal field in fact strongly modifies the conduction and valence
bands along the growth directions and  causes the separation of electrons and
holes, driving the first one towards the QD top and the latter
towards its bottom. This corresponds to
the creation of intrinsic dipoles.   If we consider two stacked dots occupied by one
exciton each, the resulting charge distribution can be seen as
two dipoles aligned along the
growth direction. This is evident in
Fig.~\ref{fig1}, where we plot the electron and hole single-particle 
 distributions
corresponding to the lowest biexcitonic state (with parallel-spin excitons)
in our GaN-based semiconductor ``macromolecule''.
The creation of stacked dipoles results in a
negative exciton-exciton coupling (or biexcitonic shift).

The theoretical approach employed to study the optical response of our GaN nanostructure is a generalization (to nitride materials) of the fully three-dimensional
exact-diagonalization scheme proposed in \onlinecite{Biolatti2}.
More specifically, we consider electrons (e) and holes (h)
confined within stacked QDs as depicted in Fig.~\ref{fig1}.
As usual, the confinement potential is modeled as parabolic in the $x-y$ plane and as a square-well potential modified by the  built-in electric field along the growth ($z$) direction.
The many-exciton optical spectra, i.e., the absorption probability
corresponding to the generic $N \to N'$ transition, is evaluated as described in \onlinecite{Biolatti2}.
In particular, here we focus on the excitonic ($0
\to 1$) and biexcitonic ($1
\to 2$) optical spectra
in the presence of the built-in electric field.
For all the structures considered, the two lowest optical
transitions correspond to the formation of direct ground-state
excitons in dot $a$ and $b$, respectively.
The biexcitonic $(1 \to 2)$ optical
spectrum
describes the creation of a second electron-hole pair in the presence
of a previously generated exciton. Here, we shall consider
parallel-spin configurations only.

Let us focus on the biexcitonic shift corresponding to the
energy difference between the ground-state biexcitonic transition (given a ground-state exciton in dot $a$)
and the ground-state excitonic transition of dot $b$.
This quantity ---a ``measure'' of
the ground-state exciton-exciton coupling--- plays a crucial role in 
all-optical quantum processors, being the key ingredient for conditional-gating schemes\cite{Biolatti1,DeRinaldis}.

Figure \ref{fig2}a shows how  the biexcitonic shift  increases
with the height of the dot. The barrier width is kept fixed and
equal to $2.5$\,nm. In curve (A) both  the height and the diameter
$D$ of the dots are varied according
 to the relation $D = 3.5L_d+3$\,nm\cite{Arley}, while
in (B) only the
height of the dot is changed.
We notice that, for realistic parameters, it is possible to achieve
 biexcitonic shifts up to $9.1$\,meV.

A few comments are now in order:
(i) When a barrier of only $2.5$\,nm separates two stacked GaAs-based
QDs, the excitonic
wavefunction is molecular-like\cite{Troiani}, forming bonding and
anti-bonding states spread over the whole macromolecule for both  electron and
hole. This effect is maximum for dots of the same size but persists even when
their dimensions are
slightly different. In GaN QDs, instead, over the range of parameters used,
the
lowest states preserve their atomic-like shape since both electron and
hole effective masses and valence/conduction-band discontinuities are
much higher than in GaAs, therefore decreasing the atomic-like wavefunction
overlap responsible for the molecular bonding.
(ii) The excitonic dipole length is roughly
proportional to the height of the dot because
of
the strong built-in electric field; therefore it is crucial to evaluate the
dependence of the exciton-exciton interaction on the  height of the QDs.
(iii) The spreading of the wavefunction affects the biexcitonic shift,
as one can notice by comparing curves A and B in Fig.~\ref{fig2}.
The biexcitonic shift is larger (up to 20\% for the parameters considered here)
when
the wavefunction is more localized, since the system is closer to the
idealized ``point-like'' particle case (see curve C in the same figure).
(iv) Our results demonstrate that there exist a wide range of parameters  for which the
biexcitonic shift is at least a few meV. This is a central prerequisite
for realizing energy-selective addressing with sub-picosecond laser pulses, as requested, for example, by all-optical
quantum information
processing schemes\cite{Biolatti1,DeRinaldis}.

Our analysis shows that the best strategy to achieve large biexcitonic shift is to grow
"high" and "small diameter"  dots. The drawback is that the oscillator
strength (OS) of
the ground-state transition strongly decreases with the height of the dot,
since it is proportional to the overlap of electron and hole wave functions.
A small value of the OS enhances the well-known difficulties of single-dot signal detection.
The lower panel in Fig.~\ref{fig2} shows that  the OS corresponding to the excitonic ground state of dot $b$ decreases super-exponentially    with the height of the dot.
As shown by the fact that curves A and B practically coincide, the height of the dot is the only parameter relevant for the OS value.
Indeed, the wavefunction spreading   due to the
width of the dot does not influence the electron-hole overlap.

In the range of height values considered in Fig.~\ref{fig2}, the OS
varies over three orders of magnitude, so care must be taken in a
future quantum information processing experiment in order to optimize at the same time biexcitonic shift and OS.
Our analysis suggests that a reasonable way to do so is to maximize
the product between the biexcitonic shift and
the logarithm of the oscillator strength. Such
quantity is
plotted in Fig.~\ref{fig3} and it is the largest for a QD height of $2.5 \div 3$\,nm.
 The curve presents a well defined maximum corresponding to
a  quantum dot  height of  2.8 nm (parabolic fit).

In conclusion we have performed a detailed investigation of exciton-exciton interaction as well as of its effect on the multi-exciton optical response in
state-of-the-art GaN-based nanostructures.
We have shown how it is possible to engineer
the interdot biexcitonic shift by varying height and width of the dots.
Our analysis provides precious indications for the realization of GaN-based quantum information processing clarifying, in particular, the crucial interplay between biexcitonic shift and  oscillator strength.

\section {Acknowledgements}
We want to thank  R. Cingolani and R. Rinaldi for  stimulating and fruitful 
discussions.
This work has been supported in part by the European Commission through
the Research Project SQID within Future and Emerging Technologies (FET)
program and in part by INFM (National Institute for the Physics of
Matter) through the Research Project PRA-SSQI

\begin{figure}
  \caption{Electron and hole particle distribution, conduction  and
valence  band structure along the growth
direction for two coupled GaN dots of, respectively, 2.5nm and 2.7 nm of
height, separated by a 2.5 nm AlN barrier. In the upper panel the dotted
line
corresponds to
hole, the solid one to the electron spatial  distribution of the
biexcitonic ground state. }
  \label{fig1} \end{figure}

\begin{figure}
  \caption{Biexcitonic shift (upper panel) and oscillator strength (lower
panel) of the
ground state transition in dot b for two
coupled GaN dots separated by a
barrier of 2.5 nm vs QD height. In curve (B) only the height of the dots
is
changed ($D= 10 nm$), while in curve (A) $D$ is varied proportionally to
the
height from 10 to 17 nm. Curve (C) shows the biexcitonic shift in the
point-
like charge approximation.
 The parameters used are : effective masses
$m_{e}
= 0.2 m_{0}$ and
$m_{h}
= m_{0}$;  in-plane parabolic confinement energy $\hbar \omega_{e} = 74 meV$
and $ \hbar \omega_{h} = 33 meV $
for
the (B) curve;  $\hbar \omega_{e} =74\div 290 meV$, $ \hbar \omega_{h} =  33\div 130
meV$ for
the (A) curve.   }
  \label{fig2} \end{figure}

\begin{figure}
  \caption{Figure of merit (biexcitonic shift times logarithm of
oscillator strength) vs QD height. The arrow indicates the maximum
obtained by a parabolic fit.       }
  \label{fig3} \end{figure}

\end{document}